\documentclass[twocolumn,english,prl,floatfix,showpacs]{revtex4-1}
\usepackage[T1]{fontenc}
\usepackage[latin9]{inputenc}
\usepackage{amsmath}
\usepackage{amssymb}
\usepackage{graphicx}



\usepackage{babel}
\begin{document}

\title{Analysis of coined quantum walks with renormalization}

\author{Stefan Boettcher and Shanshan Li}

\affiliation{Department of Physics, Emory University, Atlanta, GA 30322; USA}
\begin{abstract}
We introduce a new framework to analyze quantum algorithms with the
renormalization group (RG). To this end, we present a detailed analysis
of the real-space RG for discrete-time quantum walks on fractal networks
and show how deep insights into the analytic structure as well as
generic results about the long-time behavior can be extracted. The
RG-flow for such a walk on a dual Sierpinski gasket and a Migdal-Kadanoff
hierarchical network is obtained explicitly from elementary algebraic
manipulations, after transforming the unitary evolution equation into
Laplace space. Unlike for classical random walks, we find that the
long-time asymptotics for the quantum walk requires consideration
of a \emph{diverging} number of Laplace-poles, which we demonstrate
exactly for the closed form solution available for the walk on a \emph{1d}-loop.
In particular, we calculate the probability of the walk to overlap
with its starting position, which oscillates with a period that scales
as $N^{d_{w}^{Q}/d_{f}}$ with system size $N$. While the largest
Jacobian eigenvalue $\lambda_{1}$ of the RG-flow merely reproduces
the fractal dimension, $d_{f}=\log_{2}\lambda_{1}$, the asymptotic
analysis shows that the second Jacobian eigenvalue $\lambda_{2}$ becomes 
essential to determine the dimension of the quantum walk via $d_{w}^{Q}=\log_{2}\sqrt{\lambda_{1}\lambda_{2}}$.
We trace this fact to delicate cancellations caused by unitarity.
We obtain \emph{identical} relations for other networks, although
the details of the RG-analysis may exhibit surprisingly distinct features.
Thus, our conclusions \textendash{} which trivially reproduce those
for regular lattices with translational invariance with $d_{f}=d$
and $d_{w}^{Q}=1$ \textendash{} appear to be quite general and likely
apply to networks beyond those studied here. 
\end{abstract}
\maketitle

\section{Introduction\label{sec:Intro}}

Quantum walks~\cite{Aharonov93,Meyer96,AAKV01,Kempe03,BM_Report,PortugalBook,VA12} are rapidly achieving a central place in quantum information science. They have captured the imagination because of their wide applicability
to describe physical situations as well as computational tasks~\cite{Gro97a,Engel07,Perets08,Childs09,schreiber_2011a,Weitenberg11,Sansoni12,Schreiber12,Crespi13,PortugalBook,Manouchehri2014,Ramasesh17,Duncan17,Friedman16b}.
Such a quantum walk,  similar to random walks before them~\cite{Shlesinger84,Weiss94,Hughes96,Metzler04}, are completely described
 by the probability density
function (PDF) $\rho\left(\vec{x},t\right)$ to detect a walk at time
$t$ at site of distance $x=\left|\vec{x}\right|$ after starting
at the origin. At large times and spatial separations,
this PDF obeys the scaling collapse,
\begin{equation}
\rho\left(\vec{x},t\right)\sim t^{-\frac{d_{f}}{d_{w}}}f\left(x/t^{\frac{1}{d_{w}}}\right),\label{eq:collapse}
\end{equation}
with the scaling variable $x/t^{1/d_{w}}$, where $d_{w}$ is the
walk-dimension and $d_{f}$ is the fractal dimension of the network~\cite{Havlin87}.

The evidence that a similar scaling ansatz also describes
the behavior of quantum walks on a network is suggested by 
``weak-limit'' results that predict ballistic scaling, $d_w=1$, on $d$-dimensional
lattices~\cite{konno_2003a,grimmett_2004a}. This may be obvious for walks on a lattice in 
continuous time, which closely resemble
the tight-binding model~\cite{BM_Report,Krapivsky15}. Such a scaling is less obvious for discrete-time quantum
walks, which came to prominence as the earliest example for which
Grover's quantum search algorithms~\cite{Gro97a} can achieve a nearly quadratic
speed-up even on a square grid~\cite{AKR05,PortugalBook}.  
These require an internal
``coin'-degree of freedom to ensure unitarity, which can impact their
spreading behavior in interesting ways,  inducing localization
without disorder~\cite{Inui05,Falkner14a,schreiber_2011a,QWNComms13}.
It remains largely unexplored how the breaking of translational invariance
would affect the asymptotic scaling behavior.  That scaling as in Eq.~(\ref{eq:collapse}) 
still holds for quantum walks was argued earlier in Ref.~\cite{QWNComms13}.
We will show in the following how to \emph{analytically} determine the spreading dynamics of such a quantum walk on fractal networks. As examples, we explicitly calculate several of
the values conjectured there for the walk-dimension, in particular, on the dual Sierpinski gasket with
$d_{w}^{Q}=\log_{2}\sqrt{5}$, and on a hierarchical network (MK3) with  $d_{w}^{Q}=\log_{4}\sqrt{21}$. 
The existence and non-triviality of those values demonstrates the applicability of
Eq.~(\ref{eq:collapse}). The methods we develop to obtain it
allow the study of quantum walks in more complex environments, such
as with disorder~\cite{Maritan86,Ceccatto87} and decoherence~\cite{schreiber_2011a,kendon_2007a,romanelli_2004a}. 

\begin{figure}
\vspace{-0.2cm}

\hfill{}\includegraphics[bb=0bp 180bp 792bp 612bp,clip,width=1\columnwidth]{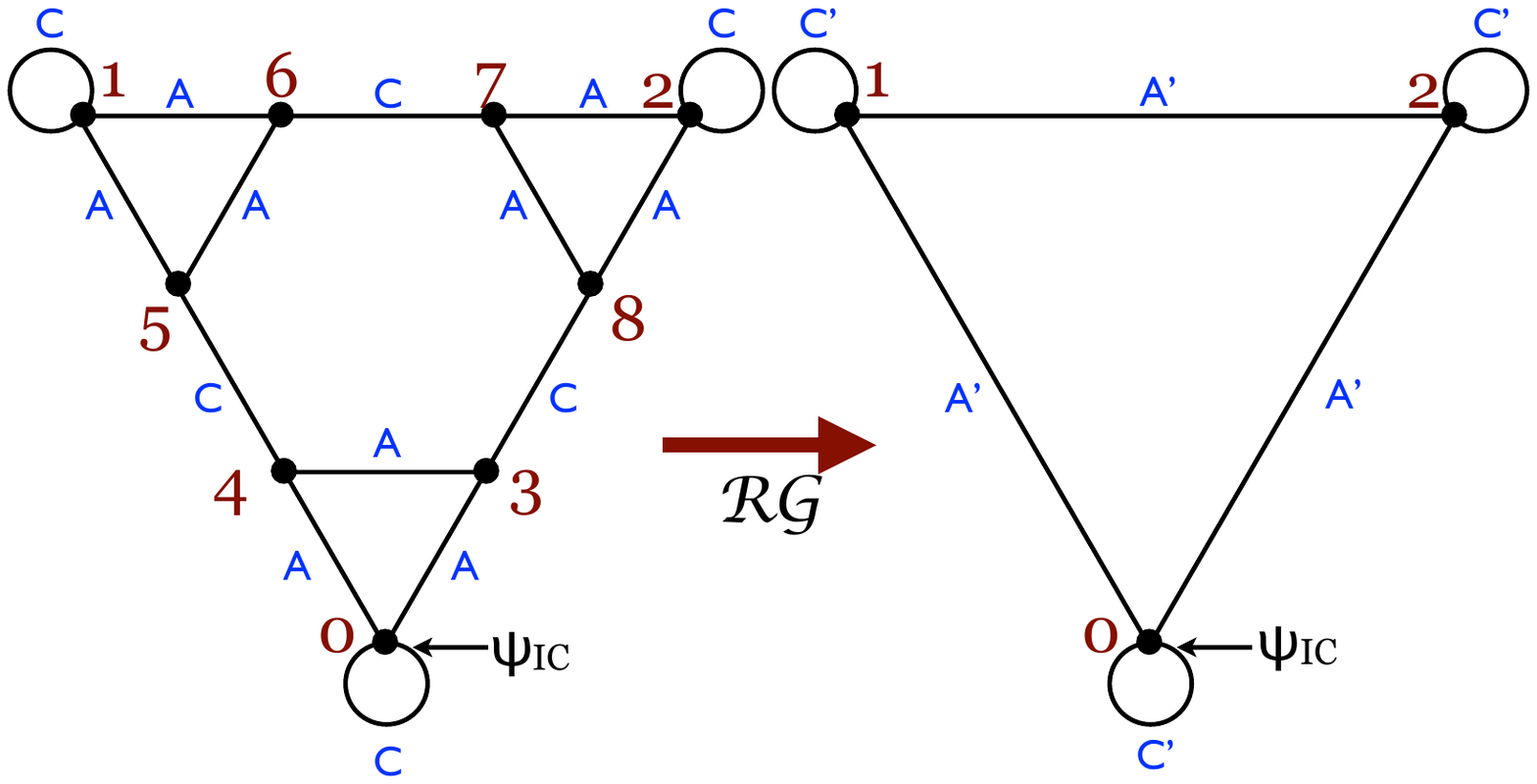}\hfill{}

\vspace{-0.3cm}
\caption{\label{fig:DSGRGstep}Depiction of the (final) RG-step in the analysis
of DSG. The letters $\left\{ A,C\right\} $ label transitions between
sites (black dots on the vertices) of the quantum walk in the form
of hopping matrices. (Only on the three outermost sites, the matrix
$C$ refers back to the same site.) Recursively, the inner-6 sites
(labeled $3,\ldots,8$) of each larger triangle (left) in DSG are
decimated to obtain a reduced triangle (right) with renormalized hopping
matrices (primed). To build a DSG of $N=3^{g}$ sites, this procedure
is applied (in reverse) $g$ times to all triangles every generation.
Each generation the base-length $L$ increases by a factor
of $b=2$, such that the fractal dimension is $d_{f}=\log_{b=2}3$.}
\end{figure}

The RG of classical random walks~\cite{Havlin87,Hughes96,Redner01}. provides a straightforward
blueprint for developing the RG for a discrete-time quantum walk,
even with the added complication of an internal coin space~\cite{Boettcher13a}.
In this way, exact RG-flow equations for quantum walks on a number
of complex networks have been derived~\cite{QWNComms13,Boettcher14b}.
Those results, for instance, have led to the conjecture that the walk
dimension $d_{w}$ in Eq.~(\ref{eq:collapse}) for a quantum walk
with a Grover coin always is \emph{half} of that for the corresponding
random walk, $d_{w}^{Q}=\frac{1}{2}\,d_{w}^{R}$~\cite{Boettcher14b}.
However, the behavior of quantum walks is complicated by the unitarity
constraint on the evolution operator. As this constraint is not necessarily
expressed by the RG recursion equations, we have argued previously~\cite{Boettcher16}
that the leading contribution in the RG-analysis had to be disregarded
to access subdominant terms. Our calculations here demonstrate the subtle and surprisingly divers ways that unitarity affects the required cancellations in the asymptotic RG-analysis.
To this end, we explicitly analyze a unitary observable, the amplitude
of a quantum walk at its starting position. As this analysis is conducted
conveniently via a Laplace-transform, a central role is assumed by
the Laplace-poles of the observable. While in the RG for classical
random walks it is usually sufficient to follow a single pole, for
the quantum walk we find it necessary to consider a\emph{ diverging}
number of such poles to facilitate a consistent evaluation. In light
of this, we show how to interpret the fixed-point of the RG-flow.
In particular, we provide an interpretation of eigenvalues of the
fixed-point Jacobian. Unlike for the classical case of a random walk~\cite{Redner01},
the RG-flow for a quantum walk must consist of at least two parameters,
to yield two relevant eigenvalues. The largest eigenvalue $\lambda_{1}$
always reflects merely the geometry of the network under consideration
by determining its fractal dimension $d_{f}$, while the dynamics
of the quantum walk in form of its walk dimension $d_{w}^{Q}$ depends
on the second eigenvalue $\lambda_{2}$, i.e., 
\begin{equation}
d_{f}=\log_{b}\lambda_{1},\qquad d_{w}^{Q}=\log_{b}\sqrt{\lambda_{1}\lambda_{2}}.\label{eq:dfw}
\end{equation}
These results are obtained for two fractal networks, the dual Sierpinski
gasket (DSG) and a Migdal-Kadanoff lattice (MK3)~\cite{Plischke94,Boettcher14b},
where each arrives at the same conclusion in quite distinct fashion,
suggesting its generality.

This paper is organized as follows: In the next Sec.~\ref{sec:Quantum-Master},
we describe the coined quantum walk, its Laplace transform, and its
implementation by example of the dual Sierpinski gasket. In Sec.~\ref{sec:Renormalization-Group-(RG)}
we derive the generic RG-recursion equations for the DSG. In Sec.~\ref{sec:The-Quantum-Walk-DSG},
we obtain the RG-flow for a specific implementation of the quantum
walk on DSG with a Grover coin. In Sec.~\ref{sec:RG-Analysis-for-the},
we analyze the fixed point of the RG-flow and apply the asymptotic
flow to determine the amplitude at the origin of the walk, with some
details of the arguments being deferred to the Appendix. In Sec.~\ref{sec:RG-Analysis-for-MK3},
we compare with the corresponding analysis on a Migdal-Kadanoff lattice.
We conclude with a discussion and outlook in Sec.~\ref{sec:Discussion}. 

\section{Quantum Master Equations\label{sec:Quantum-Master}}

The time evolution of a discrete-time quantum walk is governed by the
master equation~\cite{PortugalBook} 
\begin{equation}
\left|\Psi_{t+1}\right\rangle ={\cal U}\left|\Psi_{t}\right\rangle \label{eq:MasterEq}
\end{equation}
with unitary propagator ${\cal U}$. With $\psi_{x,t}=\left\langle x|\Psi_{t}\right\rangle $
in the $N$-dimensional site-basis $\left|x\right\rangle $ of the network,
the probability density function is given by $\rho\left(x,t\right)=\left|\psi_{x,t}^{2}\right|$.
In this basis, the propagator can be represented as a
matrix ${\cal U}_{x,y}=\left\langle x\left|{\cal U}\right|y\right\rangle$, similar to a network Laplacian, but 
with  entries that are operators in a internal coin-space which describe the transitions between
neighboring sites (``hopping matrices''). We can study the long-time
asymptotics via a discrete Laplace transform, 
\begin{equation}
\overline{\psi}_{x}\left(z\right)=\sum_{t=0}^{\infty}\psi_{x,t}z^{t},\label{eq:LaplaceT-1}
\end{equation}
as $z\to1^{-}$ implies the limit $t\to\infty$. So, Eq.~(\ref{eq:MasterEq})
becomes 
\begin{equation}
\overline{\psi}_{x}=z\sum_y{\cal U}_{x,y}\overline{\psi}_{y}+\psi_{x,t=0}.\label{eq:z_master}
\end{equation}

Due to the self-similarity of fractal networks, we can decompose ${\cal U}_{x,y}$
into its smallest sub-structure~\cite{Redner01}, exemplified by Fig.~\ref{fig:DSGRGstep}.
It shows the elementary graph-let of nine sites that is used to recursively
build the dual Sierpinski gasket (DSG) of size $N=3^{g}$ after $g$
generations. The master equations pertaining to these sites are: 
\begin{eqnarray}
\overline{\psi}_{0} & = & \left(M+C\right)\overline{\psi}_{0}+A\left(\overline{\psi}_{3}+\overline{\psi}_{4}\right)+\psi_{IC},\nonumber \\
\overline{\psi}_{\left\{ 1,2\right\} } & = & \left(M+C\right)\overline{\psi}_{\left\{ 1,2\right\} }+A\left(\overline{\psi}_{\left\{ 5,7\right\} }+\overline{\psi}_{\left\{ 6,8\right\} }\right),\nonumber \\
\overline{\psi}_{\left\{ 3,4,5,6,7,8\right\} } & = & M\overline{\psi}_{\left\{ 3,4,5,6,7,8\right\} }+C\overline{\psi}_{\left\{ 8,5,4,7,6,3\right\} }\label{eq:DSG_master}\\
 &  & \quad+A\left(\overline{\psi}_{\left\{ 0,3,1,5,2,7\right\} }+\overline{\psi}_{\left\{ 4,0,6,1,8,2\right\} }\right).\nonumber 
\end{eqnarray}
The hopping matrices $A$ and $C$ describe transitions between neighboring
sites, while $M$ permits the walker to remain on its site in a ``lazy''
walk. The inhomogeneous $\psi_{IC}$-term allows for an initial condition
$\psi_{x,t=0}=\delta_{x,0}\psi_{IC}$ for a quantum walker to start
at site $x=0$ in state $\psi_{IC}$. (It is tedious but straightforward
to generalize the following analysis to an initial condition at arbitrary
$x$ and then treat that entire section of the network accordingly.)

\section{Renormalization Group\label{sec:Renormalization-Group-(RG)}}

We now review the RG-procedure for DSG, as an illustrative example. It is identical to that discussed in Refs.~\cite{Boettcher17b,Boettcher17c}. Note that it is a vast improvement over a previous version~\cite{QWNComms13}, which assumed that the hopping matrices for each out-direction of a site            
should be distinct. However, the RG-recursions (involving five coupled nonlinear recursions with hundreds of terms each in Ref.~\cite{QWNComms13})  significantly simplify here by the assumption of symmetry, $A=B$, among the hopping matrices. As a consequence, we obtain a "lazy" walk to maintain unitarity, as we will discuss in the context of Eq.~(\ref{eq:MAC_IC}) below. 

To accomplish the decimation of the sites $\overline{\psi}_{\left\{ 3,\ldots,8\right\} }$,
as indicated in Fig.~\ref{fig:DSGRGstep}, we need to solve the linear
system in Eqs.~(\ref{eq:DSG_master}) for $\overline{\psi}_{\left\{ 0,1,2\right\} }$
 Thus, we expect that $\overline{\psi}_{\left\{ 3,\ldots,8\right\} }$
can be expressed as (symmetrized) linear combinations 
\begin{eqnarray}
\overline{\psi}_{\left\{ 3,4\right\} } & = & P\overline{\psi}_{0}+Q\overline{\psi}_{\left\{ 1,2\right\} }+R\overline{\psi}_{\left\{ 2,1\right\} },\nonumber \\
\overline{\psi}_{\left\{ 5,8\right\} } & = & R\overline{\psi}_{0}+P\overline{\psi}_{\left\{ 1,2\right\} }+Q\overline{\psi}_{\left\{ 2,1\right\} },\label{eq:Ansatz}\\
\overline{\psi}_{\left\{ 6,7\right\} } & = & Q\overline{\psi}_{0}+P\overline{\psi}_{\left\{ 1,2\right\} }+R\overline{\psi}_{\left\{ 2,1\right\} }.\nonumber 
\end{eqnarray}
Inserting this Ansatz into Eqs.~(\ref{eq:DSG_master}) and comparing
coefficients provides consistently for the unknown matrices: 
\begin{eqnarray}
P & = & \left(M+A\right)P+A+CR,\nonumber \\
Q & = & \left(M+C\right)Q+AR,\label{eq:PQRJ}\\
R & = & MR+AQ+CP.\nonumber 
\end{eqnarray}
Using the abbreviations $S=\left(\mathbb{I}-M-C\right)^{-1}A$ and
$T=\left(\mathbb{I}-M-AS\right)^{-1}C$, Eqs.~(\ref{eq:PQRJ}) have
the solution: 
\begin{eqnarray}
P&=&\left(\mathbb{I}-M-A-CT\right)^{-1}A,\nonumber\\
R&=&TP,\label{eq:solPQRJ}\\
Q&=&SR.\nonumber
\end{eqnarray}

Finally, after $\overline{\psi}_{\left\{ 3,\ldots,8\right\} }$ have
been eliminated, we find 
\begin{equation}
\overline{\psi}_{0}=\left(\left[M+2AP\right]+C\right)\overline{\psi}_{0}+A\left(Q+R\right)\left(\overline{\psi}_{1}+\overline{\psi}_{2}\right)+\psi_{IC},\label{eq:psi0RG}
\end{equation}
and similar for $\overline{\psi}_{\left\{ 1,2\right\} }$ (without
$\psi_{IC}$). By comparing coefficients between the renormalized
expression in Eq.~(\ref{eq:psi0RG}) and the corresponding, \emph{self-similar}
expression in the first line of Eqs.~(\ref{eq:DSG_master}), we can
identify the RG-recursions 
\begin{eqnarray}
M_{k+1} & = & M_{k}+2A_{k}P_{k},\nonumber \\
A_{k+1} & = & A_{k}\left(Q_{k}+R_{k}\right),\nonumber \\
C_{k+1} & = & C_{k},\label{eq:RGrecur}
\end{eqnarray}
where the subscripts refer to $k$- and $(k+1)$-renormalized forms
of the hopping matrices. These recursions evolve from the un-renormalized
($k=0$) hopping matrices with 
\begin{eqnarray}
\left\{ M,A,C\right\} _{k=0} & = & z\left\{ M,A,C\right\} .\label{eq:RGIC}
\end{eqnarray}
These RG-recursions are entirely generic and, in fact, would hold
for any walk on DSG, classical or quantum. In the following, we now
consider a specific form of a quantum walk with a Grover coin.

\section{RG-Flow for the Quantum Walk with a Grover Coin \label{sec:The-Quantum-Walk-DSG}}

To study the scaling solution for the spreading quantum walk according
to Eq.~(\ref{eq:collapse}), it is sufficient to investigate the
properties of the RG-recursion in Sec.~\ref{sec:Renormalization-Group-(RG)}
for $\left\{ M,A,C\right\} $. In the unrenormalized (``raw'') description
of the walk, these hopping matrices are chosen as 
\begin{align}
M= & \left[\begin{array}{ccc}
-\frac{1}{3} & 0 & 0\\
0 & 1 & 0\\
0 & 0 & 0
\end{array}\right]G,\label{eq:MAC_IC}\\
A= & \left[\begin{array}{ccc}
\frac{2}{3} & 0 & 0\\
0 & 0 & 0\\
0 & 0 & 0
\end{array}\right]G,\qquad C=\left[\begin{array}{ccc}
0 & 0 & 0\\
0 & 0 & 0\\
0 & 0 & 1
\end{array}\right]G.\nonumber
\end{align}
Here, we have to pay a small price for the fact that throughout, $A$
shifts weights \emph{symmetrically} to two neighboring sites within
their local triangle. The walk now must have
a ``lazy'' component, i.e., some weight may remain at each site
every update, so that $M\not=0$. Only then does the walk satisfy the unitarity conditions derived for DSG in Ref.~\cite{Boettcher17b}. The matrix $C$ shifts weight to
the one neighbor outside those triangles, as illustrated in Fig.~\ref{fig:DSGRGstep}.
These weights are the three complex components of the state vector
at each site, $\psi_{x,t}$, which are all zero at $t=0$, except
at $x=0$ where $\psi_{x=0,t=0}=\psi_{IC}$ is arbitrary but normalized,
$\left|\psi_{IC}^{2}\right|=1$. For every update, these weights are
entangled at each site before every shift via the unitary $3\times3$
coin matrix due to Grover, which is given by 
\begin{equation}
G=\frac{1}{3}\left[\begin{array}{ccc}
-1 & 2 & 2\\
2 & -1 & 2\\
2 & 2 & -1
\end{array}\right].\label{eq:GroverCoin}
\end{equation}
The walk is unitary, i.e., the norm stays preserved, because $G$
is unitary and $2A+C+M=\mathbb{I}$. Note that $G$ is also reflective,
i.e., $G^{2}=\mathbb{I}$.

Iterating the RG-recursions in Sec.~\ref{sec:Renormalization-Group-(RG)}
for the matrices in Eq.~(\ref{eq:MAC_IC}) for only $k=1$ step reveals a simple recursive pattern that suggests the Ansatz
\begin{align}
M_{k} & =\left[\begin{array}{ccc}
\frac{a_{k}}{3}-\frac{2b_{k}}{3} & 0 & 0\\
0 & z & 0\\
0 & 0 & 0
\end{array}\right]G,\label{eq:MAC_k}\\
A_{k} & =\left[\begin{array}{ccc}
\frac{a_{k}}{3}+\frac{b_{k}}{3} & 0 & 0\\
0 & 0 & 0\\
0 & 0 & 0
\end{array}\right]G,\qquad C_{k}=\left[\begin{array}{ccc}
0 & 0 & 0\\
0 & 0 & 0\\
0 & 0 & z
\end{array}\right]G.\nonumber 
\end{align}
This flow is initiated already at $k=0$ with 
\begin{equation}
a_{k=0}  =  b_{k=0}  =  z. \label{eq:abIC}
\end{equation}

Inserted into the RG-recursions in Sec.~\ref{sec:Renormalization-Group-(RG)},
these matrices exactly reproduce themselves \emph{in form} after one
iteration, $k\to k+1$, when we identify for the scalar RG-flow:
\begin{eqnarray}
a_{k+1} & = & \frac{3(3z-1)a_{k}b_{k}+(3-z)\left(a_{k}-2b_{k}\right)}{3(3-z)-(3z-1)\left(2a_{k}-b_{k}\right)},\nonumber \\
\label{eq:RGflowDSG}\\
b_{k+1} & = & \frac{\begin{array}{l}
3(3z-1)\left(3z^{2}+1\right)a_{k}b_{k}^{2}\\
\quad+2\left(3z^{3}-3z^{2}+7z-3\right)b_{k}^{2}\\
\quad-4\left(3z^{3}-6z^{2}+4z-3\right)a_{k}b_{k}\\
\quad-(3-z)\left(3+z^{2}\right)\left(a_{k}-2b_{k}\right)
\end{array}}{\begin{array}{l}
(3z-1)\left(3z^{2}+1\right)\left(2a_{k}-b_{k}\right)b_{k}\\
\quad-2\left(3z^{3}-7z^{2}+3z-3\right)a_{k}\\
\quad+4\left(3z^{3}-4z^{2}+6z-3\right)b_{k}\\
\quad+3(3-z)\left(3+z^{2}\right)
\end{array}}.\nonumber 
\end{eqnarray}
 Note that these RG-flow recursions are vastly simpler than the 5-term
recursions previously reported~\cite{QWNComms13}.

\section{RG-Analysis for the DSG\label{sec:RG-Analysis-for-the}}

We now proceed to study the fixed-point properties of the RG-flow
at $k\sim k+1\to\infty$ near $z\to1$, which builds on the discussion
in Ref.~\cite{Boettcher16}. With the choice of $a_{k}$ and $b_{k}$
in Eq.~(\ref{eq:MAC_k}), the Jacobian matrix $J=\left.\frac{\partial\left(a_{k+1},b_{k+1}\right)}{\partial\left(a_{k},b_{k}\right)}\right|_{k\to\infty}$
of the fixed point at $z=1$ and $a_{\infty}=b_{\infty}=1$
already is diagonal, with two eigenvalues, $\lambda_{1}=3$ and $\lambda_{2}=\frac{5}{3}$.
The eigenvalues correspond to those two of the five eigenvalues found
in Ref.~\cite{QWNComms13} that are relevant, i.e., they are $>1$.
Extending the expansion of Eq.~(\ref{eq:RGflowDSG}) in powers of
$\zeta=z-1$ for $k\to\infty$ to higher order, we obtain: 
\begin{eqnarray}
a_{k}\left(z\right) & \sim & 1+\zeta^{1}{\cal A}\lambda_{1}^{k}+\zeta^{2}\alpha_{k}^{(2)}+\zeta^{3}\alpha_{k}^{(3)}+\ldots,\nonumber \\
b_{k}\left(z\right) & \sim & 1+\zeta^{1}{\cal B}\lambda_{2}^{k}+\ldots,\label{eq:abFP}
\end{eqnarray}
with unknown constants ${\cal A}$ and ${\cal B}$, and with 
\begin{eqnarray}
\alpha_{k}^{(2)} & \sim & \frac{1}{2}\left({\cal A}\lambda_{1}^{k}\right)^{2}+\ldots,\label{eq:alphak}\\
\alpha_{k}^{(3)} & \sim & \frac{1}{4}\left({\cal A}\lambda_{1}^{k}\right)^{3}-\frac{1}{8}\left({\cal A}\lambda_{1}^{k}\right)^{2}\left({\cal B}\lambda_{2}^{k}\right)+\ldots,\nonumber 
\end{eqnarray}
where we have only kept leading-order terms in $k$ that contribute
in the following considerations.

For the case of a classical random walk, only the dominant eigenvalue
$\lambda_{1}$ would be relevant to determine $d_{w}^{R}=\log_{2}\lambda_{1}$
\cite{Redner01}. In contrast, in Ref.~\cite{Boettcher16} it was
conjectured that for a quantum walk the Jacobian eigenvalues provide
$d_{f}$ and $d_{w}^{Q}$ as given in Eq.~(\ref{eq:dfw}). Here,
we shall scrutinize that claim in more detail and show explicitly
how to calculate both exponents. Central to this argument is the fact
that the observable $\rho\left(x,t\right)=\left|\psi_{x,t}^{2}\right|$
in Eq.~(\ref{eq:collapse}) has Laplace-poles that only move with
$k$ on the unit circle in the complex $z$-plane, while those poles
of $a_{k}$ and $b_{k}$ move both tangentially \emph{and} radially
on the outside of that circle. That radial motion with $k$ \textendash{}
absent in $\overline{\rho}\left(x,z\right)$ \textendash{} depends
only on $\lambda_{1}$, while the tangential motion is controlled
by $\sqrt{\lambda_{1}\lambda_{2}}$. This conclusion was based on
modeling the behavior of just the two complex poles closest to $z=1$.
Although these conclusions turn out to be correct, a more detailed
analysis shows that actually $o(N)$ of such poles impinging on $z=1$
must be considered here! This we can demonstrate rigorously in Appendix~\ref{subsec:Analysis-of-the}
for the case of a quantum walk on the \emph{1d}-line. Here, we will
utilize the implications of that discussion for our analysis of DSG.

\begin{figure*}
\vspace{-0.3cm}

\hfill{}\includegraphics[bb=50bp 0bp 600bp 340bp,clip,angle=270,width=0.55\columnwidth]{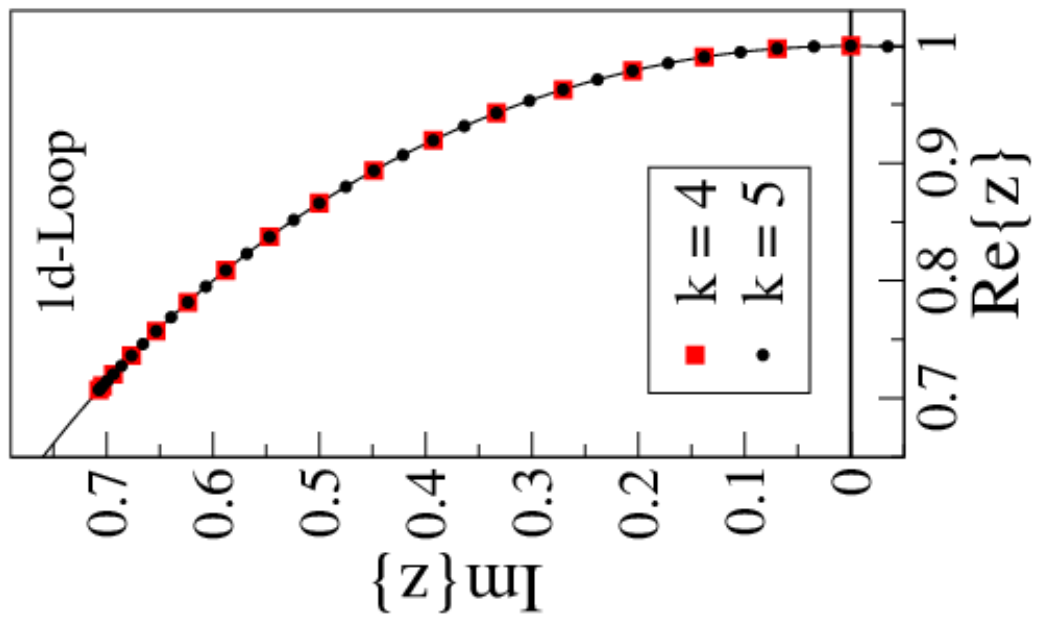}\hfill{}\includegraphics[bb=50bp 0bp 600bp 340bp,clip,angle=270,width=0.55\columnwidth]{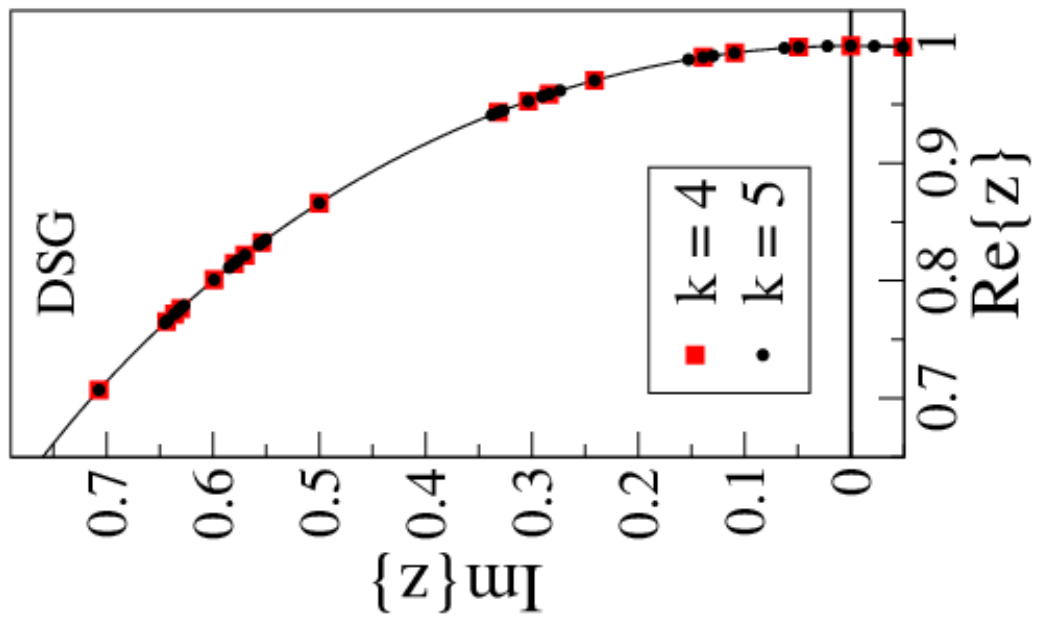}\hfill{}\includegraphics[bb=50bp 0bp 600bp 340bp,clip,angle=270,width=0.55\columnwidth]{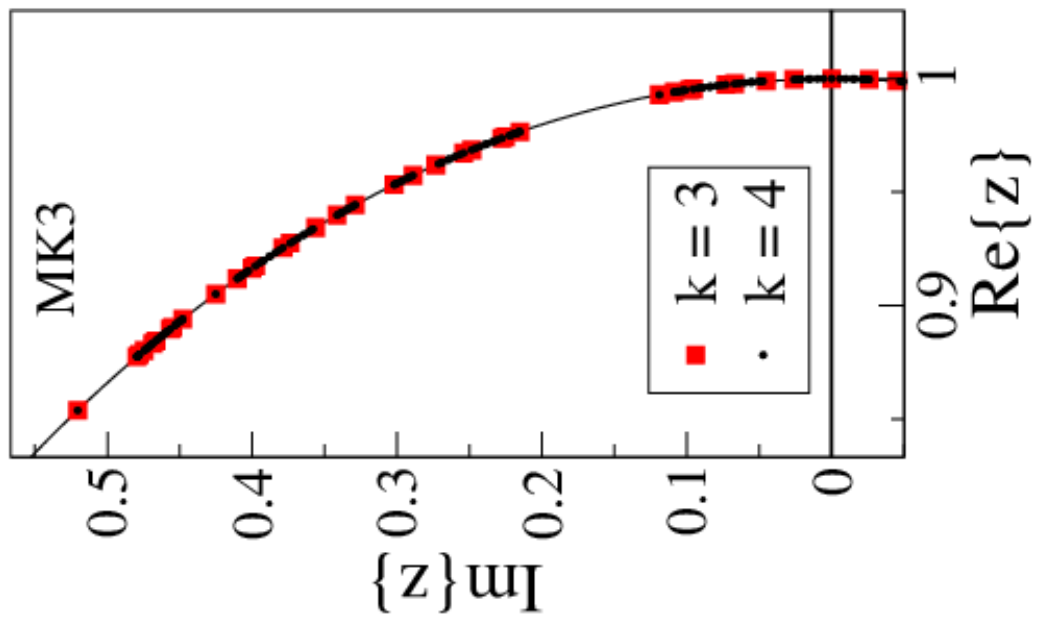}\hfill{}

\vspace{-0.3cm}
\caption{\label{fig:DSGpoles}Plot of the poles of the Laplace transforms for
the amplitude to remain at the origin, $\overline{\psi}_{0}^{(k)}(z)=X_{k}\psi_{IC}$,
in the complex-$z$ plane at two consecutive RG-steps $k$ for a quantum
walk on the 1$d$-line (left), DSG (middle), and MK3 (right). (In
these walks, poles are certain to occur in complex-conjugate pairs,
so only the upper $z$-plane is shown.) Although the pattern by which
poles evolve appears more complicated for DSG and MK3, a diverging
number of those poles progressively impinge on the real-$z$ axis
for all systems. }
\end{figure*}

Instead of $\overline{\rho}\left(x,z\right)$ in its entirety, we
focus merely on $\overline{\psi}_{0}(z)$, the amplitude at the origin
of the quantum walk on DSG. According to Fig.~\ref{fig:DSGRGstep}
and Eq.~(\ref{eq:psi0RG}), we have 
\begin{eqnarray}
\overline{\psi}_{0} & = & \left(M_{k}+C_{k}\right)\overline{\psi}_{0}+A_{k}\left(\overline{\psi}_{1}+\overline{\psi}_{2}\right)+\psi_{IC},\label{eq:DSGfinal}\\
\overline{\psi}_{\left\{ 1,2\right\} } & = & \left(M_{k}+C_{k}\right)\overline{\psi}_{\left\{ 1,2\right\} }+A_{k}\left(\overline{\psi}_{0}+\overline{\psi}_{\left\{ 2,1\right\} }\right),\nonumber 
\end{eqnarray}
which has the solution $\overline{\psi}_{0}=X_{k}\psi_{IC}$ with
\begin{equation}
X_{k}=\left[\mathbb{I}-M_{k}-C_{k}-2A_{k}\left(\mathbb{I}-M_{k}-A_{k}-C_{k}\right)^{-1}A_{k}\right]^{-1}.\label{eq:Xmatrix}
\end{equation}
Inserting Eqs.~(\ref{eq:MAC_k}) and (\ref{eq:abFP}) and expanding
(some generic component of) the matrix $X_{k}$ in powers of $\zeta=z-1$
yields 
\begin{eqnarray}
\left[X_{k}\right]_{11} & \sim & -\zeta^{-1}\frac{1}{9\left({\cal A}\lambda_{1}^{k}\right)}+\zeta^{0}\left[\frac{4}{9}+\frac{\alpha_{k}^{(2)}}{9\left({\cal A}\lambda_{1}^{k}\right)^{2}}\right]\label{eq:X11zeta}\\
 &  & +\zeta^{1}\left[\frac{\left({\cal A}\lambda_{1}^{k}\right)\alpha_{k}^{(3)}-\left(\alpha_{k}^{(2)}\right)^{2}}{9\left({\cal A}\lambda_{1}^{k}\right)^{3}}+\frac{1}{3}\left({\cal B}\lambda_{2}^{k}\right)\right]+\ldots,\nonumber \\
 & \sim & \zeta^{-1}O\left(\frac{1}{\lambda_{1}^{k}}\right)+\zeta^{0}O(1)+\zeta^{1}O\left(\lambda_{2}^{k}\right)+\ldots.\nonumber 
\end{eqnarray}
It is the cancellation of the leading $O\left(\lambda_{1}^{k}\right)$-term
at order $\zeta^{1}$ that signals the anticipated placement of the
Laplace-poles onto the unit circle in the complex $z$-plane, as demanded
by unitarity. As argued in Ref.~\cite{Boettcher16}, $\lambda_{1}$
controls the radial movement of poles with $k$ which is removed by
this cancellation, thereby exposing $\lambda_{2}^{k}$ as the relevant
contribution that controls the tangential movement of poles purely
on the unit-circle. Numerical studies of the Laplace-poles of $X_{k}$
for small values of $k$, shown in Fig.~\ref{fig:DSGpoles}, suggests
that these poles arise along arcs on the unit-circle, located symmetrically
around the real-$z$ axis due to the real-valued coin such as $G$
in Eq.~(\ref{eq:GroverCoin}), and that these poles get increasingly
dense and impinge on the real-$z$ axis at $z=1$. This picture is
borne out by our analysis of a quantum walk on the \emph{1d}-loop
in Appendix~\ref{subsec:Analysis-of-the}, which suggests the following
generalized form for the amplitude matrix at the origin: 
\begin{eqnarray}
\left[X_{k}\right]_{11} & \sim & \frac{1}{h\left(N\right)}\sum_{j=-h\left(N\right)}^{h\left(N\right)}\frac{f_{j}}{1-z\,e^{i\theta_{k}jg_{j}}},\label{eq:Xpoles}\\
 & \sim & -\zeta^{-1}\frac{f_{0}}{h\left(N\right)}\,+\zeta^{0}\frac{2S_{0}\left(N\right)}{h\left(N\right)}-\zeta^{1}\frac{4S_{2}\left(N\right)}{\theta_{k}^{2}h\left(N\right)}+\ldots,\nonumber 
\end{eqnarray}
where we defined the sums 
\begin{equation}
S_{m}\left(N\right)=\sum_{j=1}^{h\left(N\right)}\frac{f_{j}}{\left(g_{j}j\right)^{m}}.\label{eq:Sm}
\end{equation}
By analogy with the \emph{1d}-loop, we expect by the fact that the
coin in Eq.~(\ref{eq:GroverCoin}) is reflective and real that both
$f_{j}$ and $g_{j}$ are real, symmetric, and weakly-varying functions
of $j$ (but not $N$). In turn, $h\left(N\right)$ specifies how
many Laplace-poles effectively contribute to the asymptotic behavior.
If only one (or few) poles contribute, $h\left(N\right)=O(1)$, as
in the classical case~\cite{Redner01}, then $S_{m}=O(1)$ for all
$m\geq0$, and the $\zeta^{0}$-terms between Eqs.~(\ref{eq:X11zeta})
and (\ref{eq:Xpoles}) are inconsistent. The only consistent choice
entails that a diverging number of poles must be considered, $h(N)\gg1$,
specifically: $h\left(N\right)\sim\lambda_{1}^{k}=N$. This implies
that (a) $S_{0}=O(N)$ and (b) $S_{m\geq2}=O(1)$. For instance, in
the \emph{1d} quantum walk, we have $f_{j}=g_{j}=const$ such that
both (a) and (b) are satisfied. Matching the expansion also between
the $\zeta^{1}$-terms of Eqs.~(\ref{eq:X11zeta}) and (\ref{eq:Xpoles}),
we obtain $\theta_{k}^{2}h\left(N\right)\sim\lambda_{2}^{-k}$ or
$\theta_{k}^{2}\sim\lambda_{1}^{-k}\lambda_{2}^{-k}$. With $L=2^{k}$
and assuming that the scaling solution implied by Eq.~(\ref{eq:collapse})
arises via the cut-off at $\theta_{k}t\sim1$~\cite{Redner01}, i.e.,
$\theta_{k}\sim L^{-d_{w}^{Q}}$, we arrive at Eq.~(\ref{eq:dfw}).
Expanding to two more orders in powers of $\zeta$ provides further
prove of the consistency of this interpretation.

The backwards Laplace-transform of $X_{k}$ in Eq.~(\ref{eq:Xpoles})
provides for some typical component in the spinor $\psi_{0,t}$ in
a DSG of size $N=3^{k}$ that 
\begin{equation}
\psi_{0,t}\sim\frac{1}{h\left(N\right)}\sum_{j=0}^{h\left(N\right)}f_{j}\,\cos\left(\frac{jg_{j}t}{N^{d_{w}^{Q}/d_{f}}}\right).\label{eq:CosSeries}
\end{equation}
Note that due to condition (a), we have $\left|\psi_{0,t}^{(k)}\right|\sim1$
for $t=0$, as would be expected for a walk starting at $x=0$.

\section{RG-Analysis for MK3\label{sec:RG-Analysis-for-MK3}}

\begin{figure}
\hfill{}\includegraphics[bb=0bp 90bp 780bp 612bp,clip,width=1\columnwidth]{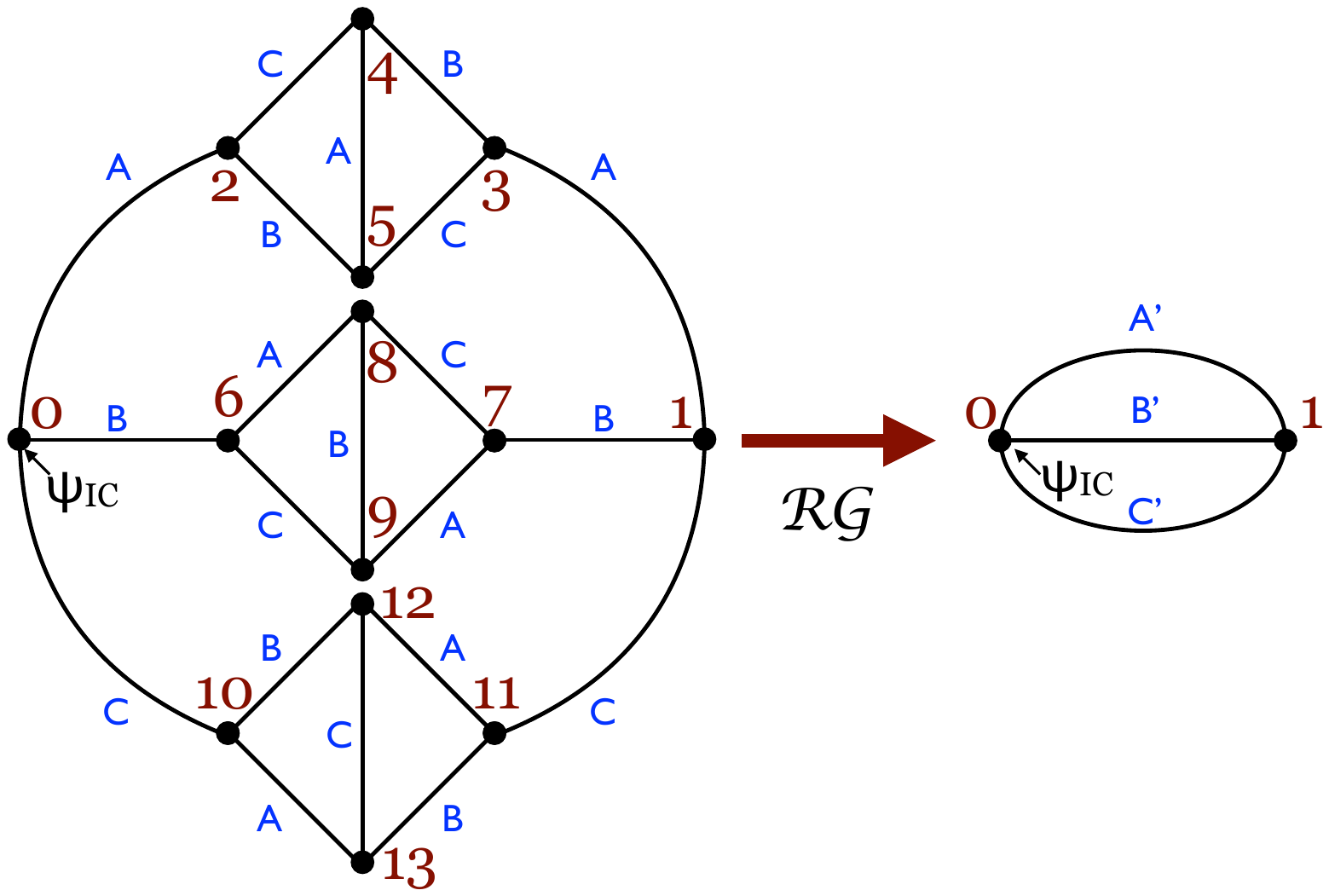}\hfill{}\caption{\label{fig:MK3RGstep}Graphic depiction of the (final) RG-step in
the analysis of MK3. The letters $\left\{ A,B,C\right\} $ label transitions
between sites (black dots on the vertices) of the quantum walk in
the form of hopping matrices. Recursively, the inner-4 sites (here
labeled $2-5,$$6-9$, and $10-13$) of each branch in MK3 are decimated
to obtain a reduced set of three lines (right) with renormalized hopping
matrices (primed). To build MK3 of $N=2\times7^{g}$ sites, this procedure
is applied in reverses $g$ times to all lines at each generation.
Note that each generation the base-length $L$ increases by a factor
of $4$, such that the fractal dimension is $d_{f}=\log_{4}7$.}
\end{figure}

To demonstrate the generality of our conclusions, we present briefly
also the corresponding analysis for another fractal network, based
on the Migdal-Kadanoff hierarchical lattices~\cite{Plischke94,Berker79}.
The RG-recursions, as depicted in Fig.~\ref{fig:MK3RGstep}, for
this case have already been presented in detail previously in Ref.~\cite{Boettcher14b}.
Again, all matrices can be parametrized with merely two scalars, most
conveniently in the form $\{A,B,C\}=\frac{a+b}{2}\left(P_{\{1,2,3\}}\cdot G\right)$
and $M=\frac{a-b}{2}\left(\mathbb{I}\cdot G\right)$, where the $3\times3$-matrices
$\left[P_{\nu}\right]_{i,j}=\delta_{i,\nu}\delta_{\nu,j}$ (with $\sum_{\nu=1}^{3}P_{\nu}=\mathbb{I}$)
facilitate the shift of the $\nu$-th component to a neighboring site.
The RG-flow was found to close for 
\begin{eqnarray}
a_{k+1} & = & \frac{\begin{array}{l}
-9a_{k}+5a_{k}^{3}+9b_{k}+3a_{k}b_{k}-17a_{k}^{2}b_{k}-3a_{k}^{3}b_{k}\\
\quad+3b_{k}^{2}+14a_{k}b_{k}^{2}-3a_{k}^{2}b_{k}^{2}-18a_{k}^{3}b_{k}^{2}
\end{array}}{\begin{array}{l}
-18-3a_{k}+14a_{k}^{2}+3a_{k}^{3}-3b_{k}-17a_{k}b_{k}\\
\quad+3a_{k}^{2}b_{k}+9a_{k}^{3}b_{k}+5b_{k}^{2}-9a_{k}^{2}b_{k}^{2}
\end{array}},\nonumber \\
b_{k+1} & = & \frac{\begin{array}{l}
-3a_{k}-a_{k}^{2}+3b_{k}+4a_{k}b_{k}-3a_{k}^{2}b_{k}\\
\quad-b_{k}^{2}+3a_{k}b_{k}^{2}+6a_{k}^{2}b_{k}^{2}
\end{array}}{\begin{array}{l}
6+3a_{k}-a_{k}^{2}-3b_{k}+4a_{k}b_{k}\\
\quad+3a_{k}^{2}b_{k}-b_{k}^{2}-3a_{k}b_{k}^{2}
\end{array}},\label{eq:rg_recursions_mk3-1}
\end{eqnarray}
with $a_{0}=b_{0}=z$. Remarkably, it can be shown that $\left|a_{k}\right|=\left|b_{k}\right|\equiv1$
for all $k$, in principle reducing the RG parameters to just two
real phases for $a_{k},b_{k}$.

Similar to DSG in Sec.~\ref{sec:RG-Analysis-for-the}, we have a
fixed point at $z=1$ with $a_{\infty}=b_{\infty}=1$. Again, the
Jacobian already is diagonal with the two eigenvalues, $\lambda_{1}=7$
and $\lambda_{2}=3$. As before, extending the expansion of Eq.~(\ref{eq:RGflowDSG})
in powers of $\zeta=z-1$ for $k\to\infty$ to higher order, we obtain:
\begin{eqnarray}
a_{k}\left(z\right) & \sim & 1+\zeta{\cal A}\lambda_{1}^{k}+\zeta^{2}\alpha_{k}^{(2)}+\zeta^{3}\alpha_{k}^{(3)}+\ldots,\nonumber \\
b_{k}\left(z\right) & \sim & 1+\zeta{\cal B}\lambda_{2}^{k}+\zeta^{2}\beta_{k}^{(2)}+\zeta^{3}\beta_{k}^{(3)}+\ldots,\label{eq:abFP_MK3}
\end{eqnarray}
with unknown constants ${\cal A}$ and ${\cal B}$, and with 
\begin{eqnarray}
\alpha_{k}^{(2)}&\sim&\frac{1}{2}\left({\cal A}\lambda_{1}^{k}\right)^{2}+\ldots,\nonumber\\
\alpha_{k}^{(3)}&\sim&\frac{1}{4}\left({\cal A}\lambda_{1}^{k}\right)^{3}+\ldots,\label{eq:alphakMK3}\\
\beta_{k}^{(2)}&\sim&\frac{1}{2}\left({\cal B}\lambda_{2}^{k}\right)^{2}+\ldots, \nonumber\\
\beta_{k}^{(3)}&\sim&-\frac{3}{80}\left({\cal A}\lambda_{1}^{k}\right)\left({\cal B}\lambda_{2}^{k}\right)^{2}+\ldots,\nonumber 
\end{eqnarray}
where we have only kept leading-order terms in $k$ relevant for the
following considerations.

The final step of the RG, shown on the right of Fig.~\ref{fig:MK3RGstep},
is given by 
\begin{eqnarray}
\overline{\psi}_{0} & = & M_{k}\overline{\psi}_{0}+\left(A_{k}+B_{k}+C_{k}\right)\overline{\psi}_{1}+\psi_{IC},\nonumber \\
\overline{\psi}_{1} & = & M_{k}\overline{\psi}_{1}+\left(A_{k}+B_{k}+C_{k}\right)\overline{\psi}_{0},\label{eq:MK3final}
\end{eqnarray}
which has the solution $\overline{\psi}_{0}=X_{k}\psi_{IC}$ with
\begin{equation}
X_{k}=\left[\mathbb{I}-M_{k}-V_{k}\left(\mathbb{I}-M_{k}\right)^{-1}V_{k}\right]^{-1},\label{eq:XmatrixMK3}
\end{equation}
abbreviating $V_{k}=A_{k}+B_{k}+C_{k}$. Inserting $\left\{ A_{k},B_{k},C_{k},M_{k}\right\} $
with the RG-flow in Eq.~(\ref{eq:abFP_MK3}) into $X_{k}$ in Eq.
(\ref{eq:XmatrixMK3}) and expanding in powers of $\zeta=z-1$ yields
for the $\left(1,1\right)$-component: 
\begin{eqnarray}
\left[X_{k}\right]_{11} & \sim & -\zeta^{-1}\left[\frac{1}{3\left({\cal B}\lambda_{2}^{k}\right)}+\frac{1}{6\left({\cal A}\lambda_{1}^{k}\right)}\right]+\zeta^{0}\left[\frac{1}{4}+\ldots\right]\nonumber \\
 &  & +\zeta^{1}\left[\frac{\left({\cal A}\lambda_{1}^{k}\right)\alpha_{k}^{(3)}-\left(\alpha_{k}^{(2)}\right)^{2}}{6\left({\cal A}\lambda_{1}^{k}\right)^{3}}-\frac{23\left({\cal A}\lambda_{1}^{k}\right)}{240}+\ldots\right]\nonumber \\
 &  & \quad+\ldots,\label{eq:X11zetaMK3}\\
 & \sim & \zeta^{-1}O\left(\frac{1}{\lambda_{2}^{k}}\right)+\zeta^{0}O(1)+\zeta^{1}O\left(\lambda_{1}^{k}\right)+\ldots.\nonumber 
\end{eqnarray}
Although the $\zeta^{1}$-term exhibits the same cancelation in the
first term as in Eq.~(\ref{eq:X11zeta}) for the DSG, other terms
of order $O\left(\lambda_{1}^{k}\right)$ remain, hiding any contributions
from $\lambda_{2}$ here. However, the highly peculiar $\zeta^{-1}$-term
\emph{also} reverses the role of the eigenvalues, now selecting $\lambda_{2}$
as the dominant contribution over $\lambda_{1}$ there. Comparison
with the expected form of $X_{k}$ in Eq.~(\ref{eq:Xpoles}) then
leads to an effective number of poles that scales sub-extensive with
the system size, $h\left(N\right)\sim\lambda_{2}^{k}=N^{\log_{7}3}$.
Consistency then also demands that $S_{0}(N)\sim h(N)$, so that $S_{m\geq2}(N)=O(1)$.
Amazingly, despite the reversal of roles between $\lambda_{1}$ and
$\lambda_{2}$, matching the $\zeta^{1}$-terms again provides $\theta_{k}^{2}\sim\lambda_{1}^{-k}\lambda_{2}^{-k}$,
which is invariant to this switch. As before, we have expanded to
two more orders in powers of $\zeta$ and found consistency throughout.
Thus, the RG of MK3  affirms the result in Eq.~(\ref{eq:dfw}),
all differences in the analysis aside.

\section{Discussion\label{sec:Discussion}}

We have provided a comprehensive description of the real-space renormalization
group treatment of discrete-time quantum walks. We have referred to
the DSG and MK3 as specific examples, but we expect that this procedure
also describes other networks. Our procedure is immediately applicable
to study the quantum search algorithm with a coin or power-law localization
in hierarchical networks, which we will present elsewhere. Especially,
our approach opens the door to a systematic consideration of universality
classes in quantum walks and search algorithm. For instance, entire
classes of coins can be studied, in particular those that might break
the symmetries that were essential do establish the current results
and the delicate cancellations these require. The methods developed here also provide the starting point for the consideration of disordered environments~\cite{Maritan86,Ceccatto87} and the discussion of localization in complex networks~\cite{QWNComms13}. Finally, approximate means can be explored on the basis of the current calculation that eventually can preserve unitarity, or allow to handle decoherence in a controlled manner that is found in any realistic implementations~\cite{kendon_2007a,Perets08,schreiber_2011a}. 

\paragraph*{Acknowledgements:}

SB acknowledges financial support from CNPq through the ``Ciência
sem Fronteiras'' program and thanks LNCC for its hospitality. SL
is  supported by the Emory University PERS program.
 \bibliographystyle{apsrev4-1}
\bibliography{../../../Boettcher}

\section{Appendix\label{sec:Appendix}}

\subsection{Analysis of the Quantum Walk on a Line\label{subsec:Analysis-of-the}}

The renormalization group treatment of the quantum walk on the 1$d$-line
\cite{Boettcher13a} provides the RG-flow 
\begin{eqnarray}
a_{k+1} & = & \frac{\sin\eta\,a_{k}^{2}}{1-2\cos\eta\,b_{k}+b_{k}^{2}},\nonumber \\
b_{k+1} & = & b_{k}+\frac{\left(b_{k}-\cos\eta\right)a_{k}^{2}}{1-2\cos\eta\,b_{k}+b_{k}^{2}}.\label{eq:1dQWrecursions1}
\end{eqnarray}
But unlike the analog expressions for DSG in Eq.~(\ref{eq:RGflowDSG})
or MK3 in Eq.~(\ref{eq:rg_recursions_mk3-1}), this RG-flow in fact
possesses a closed-form solution for all $N=2^{k}$: 
\begin{equation}
a_{k}=\frac{\cos\sigma\,\sin\eta}{\cos\left(N\nu+\sigma\right)},\quad b_{k}=\cos\eta+i\frac{\sin\left(N\nu\right)\sin\eta}{\cos\left(N\nu+\sigma\right)},\label{eq:sigmanu1d}
\end{equation}
where $\nu\left(z\right)$ and $\sigma\left(z\right)$ are determined
by matching to the initial flow, $a_{1}=z^{2}\sin\eta$ and $b_{1}=z^{2}\cos\eta$.

Previously, in Ref.~\cite{Boettcher16}, the equivalent of Eq.~(\ref{eq:Xmatrix})
for the amplitude at the starting-site of a quantum walk, $\overline{\psi}_{0}=X_{k}\psi_{IC}$,
for the \emph{1d}-line was shown to be: 
\begin{equation}
X_{k}=\left[\mathbb{I}-\left(A_{k}+B_{k}+M_{k}\right)\right]^{-1}.\label{eq:psi0in1dsearch}
\end{equation}
Here, the hopping matrices are parametrized as 
\begin{equation}
A_{k}=\left[\begin{array}{cc}
a_{k} & 0\\
0 & 0
\end{array}\right]{\cal C},\,\,B_{k}=\left[\begin{array}{cc}
0 & 0\\
0 & -a_{k}
\end{array}\right]{\cal C},\,\,M_{k}=\left[\begin{array}{cc}
0 & b_{k}\\
b_{k} & 0
\end{array}\right]{\cal C}\label{eq:PQR}
\end{equation}
after $k$ renormalization steps, with the coin matrix 
\begin{equation}
{\cal C}=\left(\begin{array}{cc}
\sin\eta & \cos\eta\\
\cos\eta & -\sin\eta
\end{array}\right).\label{eq:HadamardCoin}
\end{equation}
Eqs.~(\ref{eq:sigmanu1d}-\ref{eq:PQR}) together provide 
\begin{align}
X_{k} & =\frac{\left[\begin{array}{cc}
1-a_{k}\sin\eta-b_{k}\cos\eta & a_{k}\cos\eta-b_{k}\sin\eta\\
-a_{k}\cos\eta+b_{k}\sin\eta & 1-a_{k}\sin\eta-b_{k}\cos\eta
\end{array}\right]}{1-2a_{k}\sin\eta-2b_{k}\cos\eta+a_{k}^{2}+b_{k}^{2}},\label{eq:X1d}\\
 & =\left[\begin{array}{cc}
\frac{1}{2} & -\frac{\cot\eta}{2}\\
\frac{\cot\eta}{2} & \frac{1}{2}
\end{array}\right]+\frac{\left[\begin{array}{cc}
i\cot\eta+\sin\sigma & i-\sin\sigma\cot\eta\\
\sin\sigma\cot\eta-i & i\cot\eta+\sin\sigma
\end{array}\right]}{2\tan\frac{N\nu}{2}\cos\sigma}.\nonumber 
\end{align}
In the following, we shall express $X_{k}$ asymptotically near the
RG fixed-point for $\zeta=z-1\to0$ and $N=2^{k}\to\infty$ in \emph{three}
different ways: (1.) the exact solution, (2.) the presumed expansion
in $O\left(N\right)$ Laplace poles, and (3.) the expansion of the
RG-flow in Eq.~(\ref{eq:1dQWrecursions1}), which is typically the
only form available in non-trivial applications of the RG. The validation
of (2.) and (3.) by (1.) demonstrates our contention that, indeed,
a number of Laplace poles must be considered that diverges with $N$
to consistently interpret (3.).

\subsubsection{Exact Solution:\label{subsec:Exact-Solution:}}

We simplify matters and (w.r.o.g.) set $\eta=\frac{\pi}{4}$ in the
following. With $a_{1}=b_{1}=z^{2}/\sqrt{2}$ we find from Eq.~(\ref{eq:sigmanu1d})
that 
\begin{eqnarray}
\sin2\nu & = & i\left(\frac{1}{z^{2}}-1\right)\sqrt{1+z^{4}},\nonumber \\
\sin\sigma & = & i\,z^{2}.\label{eq:sinnusinsigma}
\end{eqnarray}
The expansion of Eq.~(\ref{eq:sigmanu1d}) in powers of $\zeta=z-1$
is now straightforward and results in 
\begin{eqnarray}
a_{k} & \sim & \zeta^{0}\frac{1}{\sqrt{2}}+\zeta^{1}\frac{N}{\sqrt{2}}+\zeta^{2}\frac{N}{2\sqrt{2}}\nonumber \\
 &  & \quad-\zeta^{3}\frac{2\left(N-2\right)\left(N-1\right)N}{3\sqrt{2}}\nonumber \\
 &  & \quad-\zeta^{4}\frac{\left(N-2\right)N\left(3N^{2}-8N+2\right)}{6\sqrt{2}}\ldots,\nonumber \\
b_{k} & \sim & \zeta^{0}\frac{1}{\sqrt{2}}+\zeta^{1}\frac{N}{\sqrt{2}}+\zeta^{2}\frac{N\left(2N-3\right)}{2\sqrt{2}}\label{eq:akbkexact}\\
 &  & \quad+\zeta^{3}\frac{\left(N-2\right)\left(N-1\right)N}{3\sqrt{2}}\nonumber \\
 &  & \quad-\zeta^{4}\frac{\left(N-2\right)N\left(4N^{2}-6N+5\right)}{12\sqrt{2}}+\ldots.\nonumber 
\end{eqnarray}
Inserted into Eq.~(\ref{eq:X1d}), we find for each component of
the matrix $X_{k}$ in Eq.~(\ref{eq:X1d}): 
\begin{eqnarray}
\left[X_{k}\right]_{11} & = & \left[X_{k}\right]_{22}\sim-\zeta^{-1}\frac{1}{N}+\zeta^{0}\frac{N-1}{2N}-\zeta^{1}\frac{2N^{2}-5}{12N}+\ldots,\nonumber \\
\left[X_{k}\right]_{12} & = & -\left[X_{k}\right]_{21}\sim\zeta^{-1}0-\zeta^{0}\frac{N-2}{2N}+\zeta^{1}0+\ldots.\label{eq:X11exact}
\end{eqnarray}
Note that a larger number of terms in Eq.~(\ref{eq:akbkexact}) is
needed that could potentially contribute to second order in Eq.~(\ref{eq:X11exact}),
due to the singular nature of $X_{k}$. However, to leading order
in $N$ in $X_{k}$, those terms finally do cancel.

\subsubsection{Expansion in Laplace-Poles:\label{subsec:Expansion-in-Laplace-Poles:}}

The long-range asymptotics (in space and time) of $X_{k}$ is determined
by its Laplace-Poles in the complex $z$-plane~\cite{Boettcher16}.
As shown in Fig.~\ref{fig:DSGpoles}, unitarity demands that these
poles are all located on the unit-circle there, and we can parametrize
them as $z_{j}=e^{i\omega_{j}}$. With $a_{1}=b_{1}=z^{2}/\sqrt{2}$
we find from Eq.~(\ref{eq:sinnusinsigma}) on the unit-circle: 
\begin{equation}
\sin\nu_{j}=-\sqrt{2}\sin\omega_{j},\quad\sin\sigma_{j}=i\,e^{2i\omega_{j}}.\label{eq:nusigma_theta}
\end{equation}
To find the Laplace-poles of $X_{k}$ in the second line of Eq.~(\ref{eq:X1d}),
we can ignore the first (non-singular) matrix and focus on the $N$-dependent
zeros of the denominator of the second, 
\begin{equation}
\nu_{j}=\frac{2\pi}{N}j\sim-\sqrt{2}\omega_{j},\qquad\left(j\in\mathbb{Z}\right),\label{eq:nupole}
\end{equation}
in accordance with Fig.~\ref{fig:DSGpoles}. Note, again, that such
a result, $\omega_{j}=j\theta_{k}$ with 
\begin{equation}
\theta_{k}=\frac{\sqrt{2}\pi}{N},\label{eq:thetak}
\end{equation}
can only be obtained because we are in possession of the closed-form
solution of the RG-flow. It will be the purpose of the next Sec.~\ref{subsec:RG-Flow-Solution:},
and of the entire RG-analysis generally, to produce the scaling of
the cut-off in time, $1/\theta_{k}$, with system size $N$.

At small $\omega_{j}$, we also have from Eq.~(\ref{eq:nusigma_theta})
that $\sin\sigma_{j}\sim i-2\omega_{j}$ and $\cos\sigma_{j}\sim\sqrt{2}$.
To evaluate the residue of $X_{k}$ at the $j^{{\rm th}}$ pole, we
obtain 
\begin{equation}
R_{j}=\lim_{z\to e^{i\omega_{j}}}\left(z-e^{i\omega_{j}}\right)X_{k}\sim-\frac{1}{N}\left[\begin{array}{cc}
1 & -i\omega_{j}\\
i\omega_{j} & 1
\end{array}\right]\sim-\frac{1}{N}\,\mathbb{I},\label{eq:residue}
\end{equation}
to leading order. Using $\omega_{j}=j\theta_{k}$, we then express
(some component of) $X_{k}$ in terms of $h(N)=O\left(N\right)$ of
such poles, 
\begin{eqnarray}
X_{k} & \sim & \sum_{j=-h(N)}^{h(N)}\,\frac{R_{j}}{z-e^{i\theta_{k}j}},\label{eq:Xpoles1d}\\
\left[X_{k}\right]_{11} & \sim & -\zeta^{-1}\frac{1}{N}-\frac{1}{N}\sum_{j=1}^{h(N)}\left[\frac{1}{\zeta+1-e^{i\theta_{k}j}}+\frac{1}{\zeta+1-e^{-i\theta_{k}j}}\right],\nonumber \\
 & \sim & -\zeta^{-1}\frac{1}{N}-\frac{1}{N}\sum_{j=1}^{h(N)}\left[\zeta^{0}+\zeta^{1}\frac{2}{\theta_{k}^{2}j^{2}}+\ldots\right],\nonumber \\
 & \sim & -\zeta^{-1}\frac{1}{N}-\zeta^{0}\frac{h(N)}{N}-\zeta^{1}\left(\sum_{j=1}^{h(N)}\frac{1}{j^{2}}\right)\frac{2}{N\theta_{k}^{2}}+\ldots.\nonumber 
\end{eqnarray}
The last line must be compared with the exact result in Eq.~(\ref{eq:X11exact}).
The first term fits exactly, and the last term does fit with the correct
choice of $\theta_{k}$ in Eq.~(\ref{eq:thetak}) and the realization
that the sum is always finite, whether $h(N)$ is small or divergent.
The key observation concerns the middle term: There, the comparison
demands that $h(N)\sim N$, i.e., that we must sum over $O\left(N\right)$
poles to make the match consistent. We thus conjecture this to be
generically true. In fact, the application of this conjecture allows
us to consistently interpret the results for DSG (and other networks).

\subsubsection{RG-Flow Solution:\label{subsec:RG-Flow-Solution:}}

Typically, such as for the case of DSG in Eq.~(\ref{eq:RGflowDSG})
or MK3 in Eq.~(\ref{eq:rg_recursions_mk3-1}), we do not possess
a closed-form solution of the RG-flow like Eq.~(\ref{eq:sigmanu1d}).
In those cases, we would proceed as in Sec.~\ref{sec:The-Quantum-Walk-DSG}
to obtain the asymptotic expansion of the RG-flow by expanding around
the fixed point at $z=1$. This expansion~\cite{Boettcher13a} finds
the Jacobian eigenvalues $\lambda_{1}=\lambda_{2}=2$ to first order
and continues to yield: 
\begin{eqnarray}
a_{k}\left(z\right) & \sim & \frac{1}{\sqrt{2}}+\zeta{\cal A}\lambda_{1}^{k}+\zeta^{2}\alpha_{k}^{(2)}+\zeta^{3}\alpha_{k}^{(3)}+\ldots,\label{eq:akbkFP}\\
b_{k}\left(z\right) & \sim & \frac{1}{\sqrt{2}}+\zeta{\cal B}\lambda_{2}^{k}+\zeta^{2}\beta_{k}^{(2)}+\zeta^{3}\beta_{k}^{(3)}+\ldots,\nonumber 
\end{eqnarray}
with 
\begin{eqnarray}
\alpha_{k}^{(2)} & \sim & \frac{1}{\sqrt{2}}\left({\cal A}\lambda_{1}^{k}\right)^{2}-\frac{1}{\sqrt{2}}\left({\cal B}\lambda_{2}^{k}\right)^{2}+\ldots,\nonumber \\
\alpha_{k}^{(3)} & \sim & \frac{1}{3}\left({\cal A}\lambda_{1}^{k}\right)^{3}-\frac{5}{3}\left({\cal A}\lambda_{1}^{k}\right)\left({\cal B}\lambda_{2}^{k}\right)^{2}+\ldots,\nonumber \\
\beta_{k}^{(2)} & \sim & \sqrt{2}\left({\cal A}\lambda_{1}^{k}\right)\left({\cal B}\lambda_{2}^{k}\right)+\ldots,\\
\beta_{k}^{(3)} & \sim & \frac{4}{3}\left({\cal A}\lambda_{1}^{k}\right)^{2}\left({\cal B}\lambda_{2}^{k}\right)-\frac{2}{3}\left({\cal B}\lambda_{2}^{k}\right)^{3}+\ldots.\nonumber 
\end{eqnarray}
Inserting Eq.~(\ref{eq:akbkFP}) into the first line of Eq.~(\ref{eq:X1d})
yields: 
\begin{eqnarray}
\left[X_{k}\right]_{11}=\left[X_{k}\right]_{22} & \sim & -\zeta^{-1}\frac{{\cal A}+{\cal B}}{\sqrt{2}\,\lambda_{1,2}^{k}\left({\cal A}^{2}+{\cal B}^{2}\right)}+\zeta^{0}\frac{1}{2}\nonumber \\
 &  & \quad-\zeta^{1}\frac{\lambda_{1,2}^{k}\left({\cal A}+{\cal B}\right)}{6\sqrt{2}}+\ldots,\label{eq:X11RG1d}\\
\left[X_{k}\right]_{12}=-\left[X_{k}\right]_{21} & \sim & \zeta^{-1}\frac{{\cal A}-{\cal B}}{\sqrt{2}\,\lambda_{1,2}^{k}\left({\cal A}^{2}+{\cal B}^{2}\right)}-\zeta^{0}\frac{1}{2}\nonumber \\
 &  & \quad+\zeta^{1}\frac{\lambda_{1,2}^{k}\left({\cal A}-{\cal B}\right)}{6\sqrt{2}}+\ldots,\nonumber 
\end{eqnarray}
where we have kept only terms to leading order in large $\lambda_{1,2}^{k}$
for each order of $\zeta$. With the (global) exact solution in Eq.
(\ref{eq:akbkexact}), we can easily identify ${\cal A}={\cal B}=\frac{1}{\sqrt{2}}$,
however, a (local) asymptotic analysis does not provide such information.
Thus, we would not realize the accidental cancelation of the $\zeta^{\pm1}$-terms
in the off-diagonal elements of $X_{k}$ in Eq.~(\ref{eq:X11RG1d}).
As those terms are appearing only as divergent as the ones on the
diagonal, it will not affect the conclusions.

In summary, RG would tell us that each component of $X_{k}$ has the
form 
\begin{equation}
\left[X_{k}\right]_{ij}\sim\zeta^{-1}O\left(\lambda_{1,2}^{-k}\right)+\zeta^{0}O(1)+\zeta^{1}O\left(\lambda_{1,2}^{k}\right)+\ldots.\label{eq:Xk_generalRG}
\end{equation}
Thus, by comparing the $\zeta^{-1}$-term between Eq.~(\ref{eq:Xk_generalRG})
and the expected form of the amplitude in Eq.~(\ref{eq:Xpoles1d}),
we determine $h(N)\sim\lambda_{1,2}^{k}=N$. This allows us to conclude
that $\log_{2}\lambda_{1}=d_{f}$, based on the fact that this relation
has been observed on all networks studied thus far. This relation
may seem obvious from $\lambda_{1}^{k}=2^{k}$ but could well have
be a mere coincidence. (For example, it would be wrong to conclude
generally that $\log_{2}\lambda_{2}$ provides $d_{f}$!) Furthermore,
by comparing the $\zeta^{1}$-terms provides that $\lambda_{1,2}^{k}\sim1/\left(N\theta_{k}^{2}\right)$,
i.e, that the temporal cut-off scales as $\theta_{k}\sim\lambda_{1,2}^{-k}\sim1/N$,
which implies by Eq.~(\ref{eq:collapse}) that $d_{w}^{Q}=\log_{2}\lambda_{1,2}=1$.
Note, though, that our main conclusion here is that by comparing order
$\zeta^{0}$-terms we \emph{must} assume $h(N)\sim N$ for a consistent
interpretation, i.e., $O(N)$ such poles contribute to this result
to make Eq.~(\ref{eq:Xk_generalRG}) consistent with the corresponding
expansion of Laplace-poles in Eq.~(\ref{eq:Xpoles1d}). Luckily,
we do not need to know anything \emph{about} those poles. 
\end{document}